\documentclass[aps,prb,twocolumn,amsmath,amssymb,showpacs,superscriptaddress]{revtex4-1}

\usepackage{graphicx}
\usepackage{amsmath,amssymb,amsfonts}
\usepackage{textcomp}
\usepackage{hyperref}
\usepackage{gensymb}
\usepackage{verbatim}
\usepackage{amsmath}
\hypersetup{breaklinks=true,colorlinks=true,urlcolor=black}
\usepackage{color}
\usepackage{soul}
\DeclareGraphicsExtensions{.jpg,.pdf,.png,.eps}

\begin{document}
\title{Growth and characterization of BaZnGa}

\author{Na Hyun Jo}
\affiliation{Ames Laboratory, U. S. DOE, Iowa State University, Ames, Iowa 50011, USA}
\affiliation{Department of Physics and Astronomy, Iowa State University, Ames, Iowa 50011, USA}

\author{Qisheng Lin}
\affiliation{Ames Laboratory, U. S. DOE, Iowa State University, Ames, Iowa 50011, USA}
\affiliation{Department of Chemistry, Iowa State University, Ames, Iowa 50011, USA}

\author{Manh Cuong Nguyen}
\affiliation{Ames Laboratory, U. S. DOE, Iowa State University, Ames, Iowa 50011, USA}
\affiliation{Department of Physics and Astronomy, Iowa State University, Ames, Iowa 50011, USA}

\author{Udhara S. Kaluarachchi}
\affiliation{Ames Laboratory, U. S. DOE, Iowa State University, Ames, Iowa 50011, USA}
\affiliation{Department of Physics and Astronomy, Iowa State University, Ames, Iowa 50011, USA}

\author{William. R. Meier}
\affiliation{Ames Laboratory, U. S. DOE, Iowa State University, Ames, Iowa 50011, USA}
\affiliation{Department of Physics and Astronomy, Iowa State University, Ames, Iowa 50011, USA}

\author{Soham Manni}
\affiliation{Ames Laboratory, U. S. DOE, Iowa State University, Ames, Iowa 50011, USA}
\affiliation{Department of Physics and Astronomy, Iowa State University, Ames, Iowa 50011, USA}

\author{Savannah S. Downing}
\affiliation{Ames Laboratory, U. S. DOE, Iowa State University, Ames, Iowa 50011, USA}
\affiliation{Department of Physics and Astronomy, Iowa State University, Ames, Iowa 50011, USA}

\author{Anna E. B\"ohmer}
\affiliation{Ames Laboratory, U. S. DOE, Iowa State University, Ames, Iowa 50011, USA}
\affiliation{Department of Physics and Astronomy, Iowa State University, Ames, Iowa 50011, USA}

\author{Tai Kong}
\affiliation{Ames Laboratory, U. S. DOE, Iowa State University, Ames, Iowa 50011, USA}
\affiliation{Department of Physics and Astronomy, Iowa State University, Ames, Iowa 50011, USA}

\author{Yang Sun}
\affiliation{Ames Laboratory, U. S. DOE, Iowa State University, Ames, Iowa 50011, USA}
\affiliation{Department of Physics and Astronomy, Iowa State University, Ames, Iowa 50011, USA}

\author{Valentin Taufour}
\affiliation{Ames Laboratory, U. S. DOE, Iowa State University, Ames, Iowa 50011, USA}
\affiliation{Department of Physics and Astronomy, Iowa State University, Ames, Iowa 50011, USA}

\author{Cai-Zhuang Wang}
\affiliation{Ames Laboratory, U. S. DOE, Iowa State University, Ames, Iowa 50011, USA}
\affiliation{Department of Physics and Astronomy, Iowa State University, Ames, Iowa 50011, USA}

\author{Kai-Ming Ho}
\affiliation{Ames Laboratory, U. S. DOE, Iowa State University, Ames, Iowa 50011, USA}
\affiliation{Department of Physics and Astronomy, Iowa State University, Ames, Iowa 50011, USA}

\author{Sergey L. Bud'ko}
\affiliation{Ames Laboratory, U. S. DOE, Iowa State University, Ames, Iowa 50011, USA}
\affiliation{Department of Physics and Astronomy, Iowa State University, Ames, Iowa 50011, USA}

\author{Paul C. Canfield}\email[]{canfield@ameslab.gov}
\affiliation{Ames Laboratory, U. S. DOE, Iowa State University, Ames, Iowa 50011, USA}
\affiliation{Department of Physics and Astronomy, Iowa State University, Ames, Iowa 50011, USA}

\begin{abstract}
We report the growth, structure and characterization of BaZnGa, identifying it as the sole known ternary compound in the Ba-Zn-Ga system. Single crystals of BaZnGa can be grown out of excess Ba-Zn and adopt a tI36 structure type. There are three unique Ba sites and three M\,=\,Zn/Ga sites. Using DFT calculations we can argue that whereas one of these three M sites is probably solely occupied by Ga, the other two M sites, most likely, have mixed Zn/Ga occupancy. Temperature dependent resistivity and magnetization measurements suggest that BaZnGa is a poor metal with no electronic or magnetic phase transitions between 1.8\,K and 300\,K. 
\end{abstract}

\maketitle 

\section{Introduction}
Poorly explored binary and ternary phase diagrams offer promising regions for the discovery of new materials. Over the past several years, exploration of the Pt-Sn, Sc-Zn and R-Cd (R = rare earth) binary phase diagrams has led to the discovery of novel properties in PtSn$_4$,\cite{Mun2012,WuPtSn2016} as well as the discovery of i-ScZn\cite{Canfield20102} and i-RCd\cite{Goldman2013} icosahedral quasicrystals. Similar exploration of the R-Fe-Zn and Fe-Li-N ternary phase spaces led to finding exceptional correlated electron effects in the RFe$_2$Zn$_{20}$ materials\cite{Jia20072,torikachvili2007} and  Fe-based local moment anisotropy comparable or greater than that found in rare earth ions.\cite{Jesche2014} In most of these cases, the ability to work with volatile and/or reactive materials was a key step to being able to access the requisite phase space.  Over the past several years we have mastered the handling and use of both Ba (reactive) and Zn (volatile). This opens the exploration of hither-to poorly explored, ternary Ba-Zn-X phase spaces.  Herein, we report the discovery and characterization of the only known Ba-Zn-Ga ternary compound, BaZn$_{1-\delta}$Ga$_{1+\delta}$ or BaZnGa for short.
 
\section{Experiment}

\begin{figure}
	\includegraphics[scale=1]{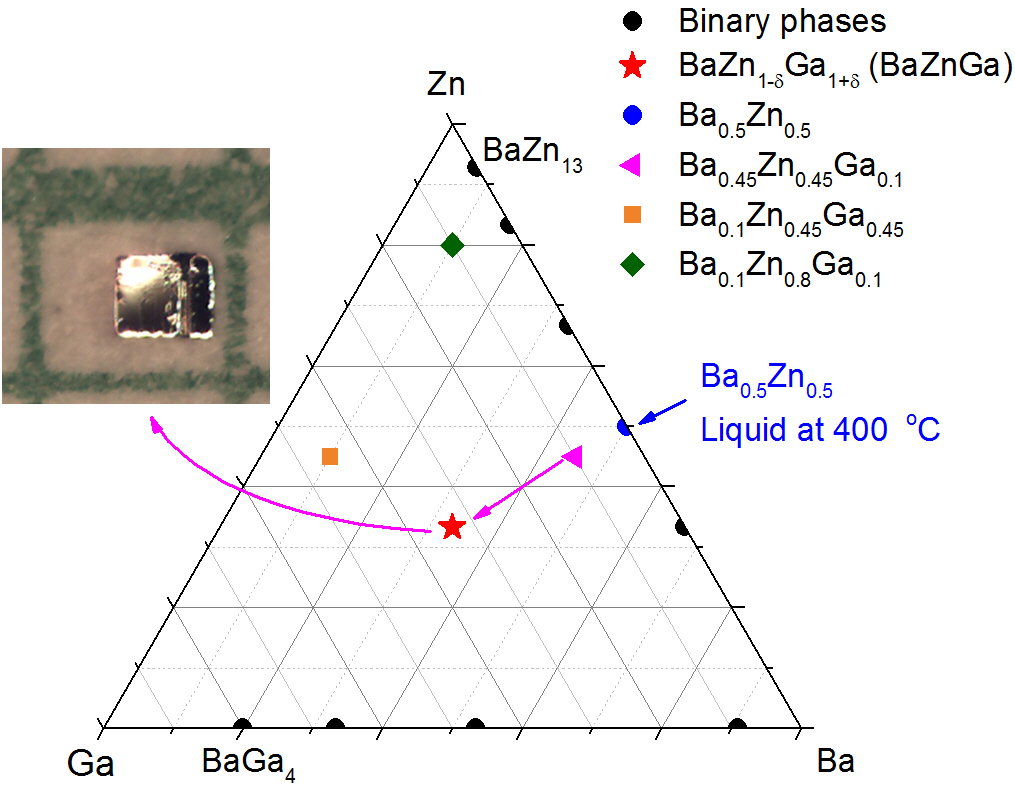}%
	\caption{(color online) Ternary phase diagram of Ba-Zn-Ga. Photograph of a single crystal of BaZnGa on a mm grid.
		\label{picture}}
\end{figure}

Single crystals of BaZnGa were grown out of a low temperature Ba-Zn eutectic.\cite{ASM2000}
In order to check the utility of the eutectic and possible reaction between alumina crucible and Ba, we put Ba$_{0.5}$Zn$_{0.5}$ into an alumina crucible and sealed in an amorphous silica tube. We heated up to 800\,\celsius\, over 5 hours, kept it there for 12 hours, cooled down to 400\,\celsius\, over 12 hours, and then finally decanted using a centrifuge.\cite{Canfield1992}
All of the material was in a liquid state and passed through an alumina frit.\cite{Canfield2016} In addition, there was no sign of attack on the alumina crucible by Ba or Zn. 

As a next step, Ba, Zn and Ga were placed into a new alumina crucible with an initial stoichiometry of Ba$_{0.45}$Zn$_{0.45}$Ga$_{0.1}$ (shown on the ternary phase diagram in Fig.\ref{picture}) and sealed in an amorphous silica tube. The ampule was heated up to 800\,\celsius\, over 5 hours, kept there for 5 hours, cooled down to 400\,\celsius\, over 60 hours, and then finally decanted using a centrifuge. The obtained crystals are plate-like (Fig. \ref{picture}). The crystals are sensitive to air, so all the sample preparations and handling were done inside a glove-box filled with argon or nitrogen. We also tried two other initial compositions, Ba$_{0.1}$Zn$_{0.45}$Ga$_{0.45}$ and Ba$_{0.1}$Zn$_{0.8}$Ga$_{0.1}$. These are also shown on Fig.\ref{picture}, but neither resulted in any ternary phase growth. Instead they resulted primarily in binary phases of BaGa$_4$, and BaZn$_{13}$, respectively.

Single crystal X-ray diffraction intensity data for BaZnGa were collected at room temperature (293\,K) using a Bruker SMART APEX II diffractometer (Mo\,K$_{\alpha}$ radiation, $\lambda$\,=\,0.71073 $\textrm{\AA}$). (BaZnGa single crystals were examined under microscopy and sealed into capillaries in a glove box.) Data reduction, integration, unit cell refinements, and absorption corrections were done with the aid of programs integrated into the APEX2 software package.\cite{Bruker2002, Bruker2013} Space group determination, Fourier Synthesis, and full-matrix least-squares refinements on $F^2$ were carried out with SHELXTL 6.1.\cite{Bruker2000} The program Superflip\cite{Palatinus2007} was used to yield the initial structural model, in which three independent sites could be assigned to Ba and the remaining three to Ga (or Zn) based on peak intensity and interatomic distances. The occupancy parameter for each site was found to be close to full occupancy ($<$ 1 $\sigma$) when separate refinements with unfixed occupancy parameter were tried. However, attempts to refine the Zn/Ga mixture on these three Ga (or Zn) sites were not successful, which is common for systems containing elements with similar scattering powers. Table I gives the crystal data and structure refinement for BaZnGa and Table II lists the refined atomic positions and equivalent isotropic displacement parameters.

Magnetization measurements were performed in a Quantum Design, Magnetic Property Measurement System (MPMS), SQUID magnetometer for $H$\,=\,10\,kOe and 1.8\,K$\le\,T\,\le\,300$\,K. The samples for the magnetization were put into a gel capsule without a well defined orientation. The magnetization data of BaZnGa were obtained by subtracting the background signal of the empty gel capsule from the sample plus gel capule data set. Samples for $ac$ resistivity measurement were prepared in a standard 4-probe geometry. Au wires (12.7\,\micro m diameter) were attached to the crystal using DuPont 4929N silver paint inside the glove box as shown in the inset to Fig.\,\ref{property} below. The temperature dependent electrical transport measurements were carried out in a Quantum Design Physical Property Measurement System (PPMS) for $1.8\,\textrm K\,\leq\,T\,\leq\,300\,\textrm K$. 

 Elemental analysis was done via Oxford Aztec energy dispersive spectroscopy(EDS) using FEI quanta 250 FEG scanning electron microscope. 

\section{Results and Analysis}

\subsection{\texorpdfstring{Crystal structure and composition}{space}}

\begin{table}[]
	\centering
	\caption{Crystal data and structure refinement for BaZnGa.}
	\label{table1}
	\begin{tabular}{lllll}
		\hline \hline
		Empirical formula                        & BaZnGa   &  &  &  \\ \hline
		Space group                              & $I4/mmm$ (tI36)            &  &  &  \\
		Unit cell dimensions                     & $a$ = 6.3652(2) $\AA$     &  &  &  \\
		                                         & $c$ = 25.593(1) $\AA$     &  &  &  \\
		Volume ($\AA^3$)                         & 1036.91(8) $\AA^3$        &  &  &  \\
		Z                                        & 12                        &  &  &  \\
		Reflections collected                    & 8645 [R$_{int}$ = 0.0557] &  &  &  \\
		Data / restraints / parameters           & 557 / 0 / 21              &  &  &  \\
		Goodness-of-fit on F$^2$                 & 1.163                     &  &  &  \\
		Final R indices [I$\textgreater$ 2$\sigma (I)$] & R1 = 0.0462, wR2 = 0.1036 &  &  &  \\
		R indices (all data)                     & R1 = 0.0480, wR2 = 0.1043 &  &  &  \\
		Largest diff. peak and hole              & 3.933 and -2.128 e.$\AA^{-3}$ &  &  &  \\ \hline \hline
	\end{tabular}
\end{table}

\begin{table}[]
	\centering
	\caption{The refined atomic positions and equivalent isotropic displacement parameters for BaZnGa.}
	\label{table2}
	\begin{tabular}{lllllll}
		\hline \hline
		Atom   & Wyck. & Site Symm. & X         & y & z         & U$_{eq}$ ($\AA^2$) \\ \hline 
		Ba1    & 4d    & -4m2       & 1/2       & 0   & 1/4        & 0.015(1) \\
		Ba2    & 4e    & 4mm        & 1/2       & 1/2 & 0.1322(1) & 0.041(1) \\
		Ba3    & 4c    & mmm        & 1/2       & 0   & 0         & 0.046(1) \\ 
		M1     & 4e    & 4mm        & 0         & 0   & 0.0508(1) & 0.015(1) \\
		M2     & 4e    & 4mm        & 0         & 0   & 0.1987(1) & 0.015(1) \\
		M3     & 16n   & .m.        & 0.2955(2) & 0   & 0.1236(1) & 0.015(1) \\ \hline \hline	        
	\end{tabular}
\end{table}

\begin{figure}
	\includegraphics[scale=1.2]{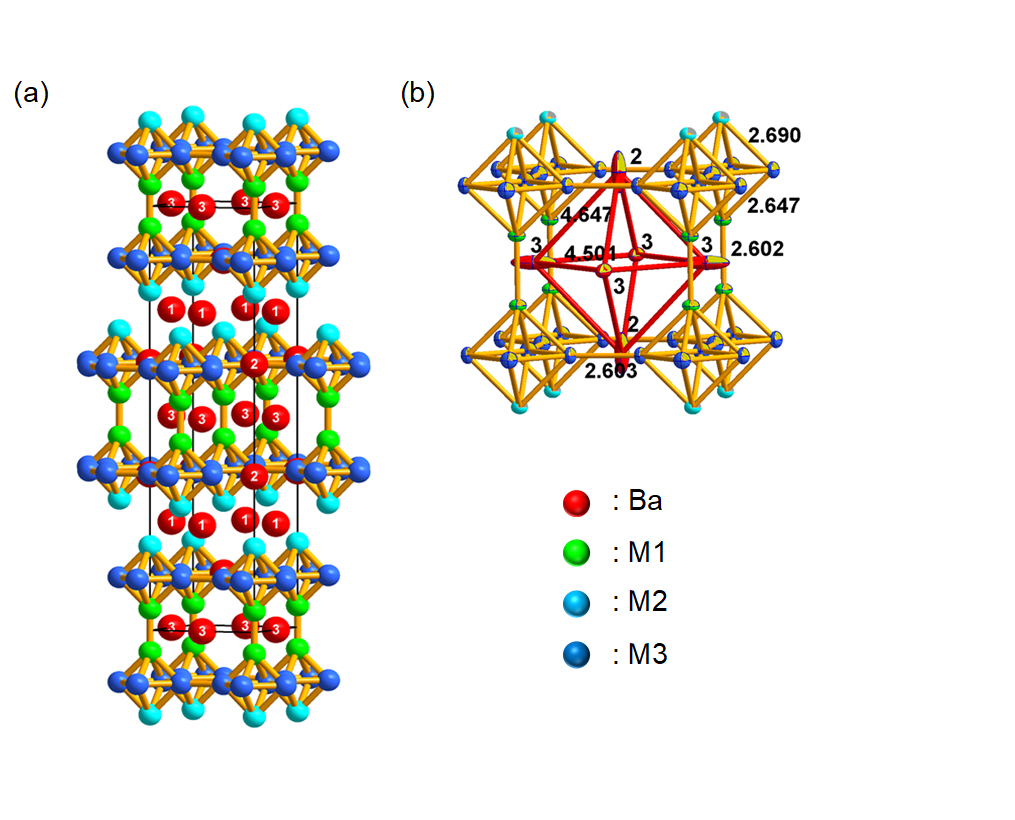}%
	\caption{(color online) (a) The structure of BaZnGa at a unit cell level, containing two identical double octahedral layers. Red spheres denote Ba atoms, with numbers showing independent atoms. Green, cyan, and blue spheres denote atoms on the M1, M2, and M3 sites, respectively. (b) The repeating structural unit in one double octahedral layer. Important Ba-Ba and (Zn/Ga)-(Zn/Ga) bond distances are marked. 
		\label{structure}}
\end{figure}

Based on our growth and identification, BaZnGa is the only, currently known, ternary phase in the Ba-Zn-Ga system. It crystallizes in a new structural type, with Pearson symbol of $tI$36. This is distinct from both BaZn$_2$ (KHg$_2$ type, oI12) and BaGa$_2$ (AlB$_2$ type, hP3) binary compounds. The structure features a packing of double octahedral layers of Zn and Ga at $z$\,=\,0 and 1/2. These layers are identical but shifted by half unit cell in both the $a$ and $b$ directions and separated by a layer of Ba1 atoms (Fig.\,\ref{structure} (a)). The Ba2 and Ba3 atoms are located at vacancy centers of four neighboring octahedra, and they form octahedra slightly elongated along the $c$ axis, as indicated by the interatomic distance of Ba3\,-\,Ba3 (4.501\,$\AA$) and Ba2\,-\,Ba3 (4.647\,$\AA$). In each double octahedral layer, neighboring (Zn/Ga) octahedra are connected by M3\,-\,M3 inter-cluster bonds in the $ab$ plane and M1\,-\,M1 bonds along $c$; both are $\sim$\,2.60\,$\AA$. In contrast, the intra-cluster bond distances of the octahedra are longer (2.65\,$\AA$\,-\,2.69\,$\AA$), suggesting more delocalized bonding interactions. Crystal orbital Hamilton Population (COHP) analyses reveal that the inter-cluster bonding interactions are about 2.23\,-\,2.28\,eV/bond.mol, i.e. $\sim$\,30\,$\%$ larger than those of intra-cluster bonds (1.65\,-\,1.75\,eV/bond.mol). The arrangement between the Ba$_{6}$ octahedron and its neighboring octahedra (Fig.\,\ref{structure} (b)) can be viewed as an inversed Perovskite structure with a general formula of ABX$_3$, in which the large size cation A is now replaced by the Zn/Ga octahedra and the anion X by Ba, whereas the small size cation B is missing. 

Although analysis of the X-ray data cannot shed light onto the Zn/Ga ratio, we were able to perform semi-quantitative composition analysis via EDS. The Ba:Zn:Ga ratio was consistent with 1\,:\,1\,:\,1, although it did suggest that there may be more Ga than Zn, i.e. BaZn$_{1-\delta}$Ga$_{1+\delta}$.

\subsection{\texorpdfstring{Density Functional Theory}{space}}

Given that the x-ray scattering strength of Zn and Ga are very close, the occupations of Zn and Ga on 3 Wyckoff sites (two 4e-sites, namely M1 and M2, and 16n-site, namely M3) cannot be resolved by single crystal analysis. First-principles density functional theory (DFT)\cite{Kohn1965} calculations were performed to investigate the occupancy of Zn and Ga on these 3 sites. In order to allow for an unequal amount of Zn and Ga, one thousand different structures with 8 Zn and 16 Ga atoms scattered randomly on those 3 sites were generated and their corresponding energies were calculated by DFT. The DFT calculations were performed using Vienna $Ab-Initio$ Simulation Package (VASP)\cite{Kresse1996} with projector-augmented wave pseudopotential method \cite{Blochl1994, Kresse1999} and plane wave basis. The generalized-gradient approximation parameterized by Perdew, Burke, and Ernzerhof\cite{Perdew1996} was used for the exchange correlation energy functional. The energy cutoff was 360\,eV and the Monkhorst-Pack's scheme\cite{Monkhorst1976} was used for Brillouin zone sampling with a k-point mesh of 6\,$\times$\,6\,$\times$\,2. All low energy structures, with energy window of 0.25\,eV/unit cell, which is corresponding to thermal energy of liquid Nitrogen (77\,K), were collected for site occupation analysis. We find that the M2 site is fully occupied by Ga in all low-energy structures. The M1 site and M3 site are occupied by both Zn and Ga with average occupations Zn$_{1.85}$Ga$_{2.15}$ and Zn$_{6.15}$Ga$_{9.85}$, respectively, which are very close to the totally random occupation of Zn and Ga on these 2 sites: Zn$_{1.60}$Ga$_{2.40}$ and Zn$_{6.40}$Ga$_{9.60}$. Therefore, we conclude that the M2 site is likely occupied by Ga only, whereas and M1 and M3 sites are occupied by Ga and Zn randomly. We would like to note here that, amongst the M1, M2 and M3 sites, M2 site has a coordination number of 4, whereas both M1 and M3 have coordination numbers of 5. This distinction of coordination number could be the driving force for the distinct occupation of the M2 site. The very strong predisposition of Ga to occupy the M2 site shown above implies that the M2 site is primarily or fully occupied by Ga in the BaZnGa ternary.

\subsection{\texorpdfstring{Physical properties}{space}}

\begin{figure}
	\includegraphics[scale=1]{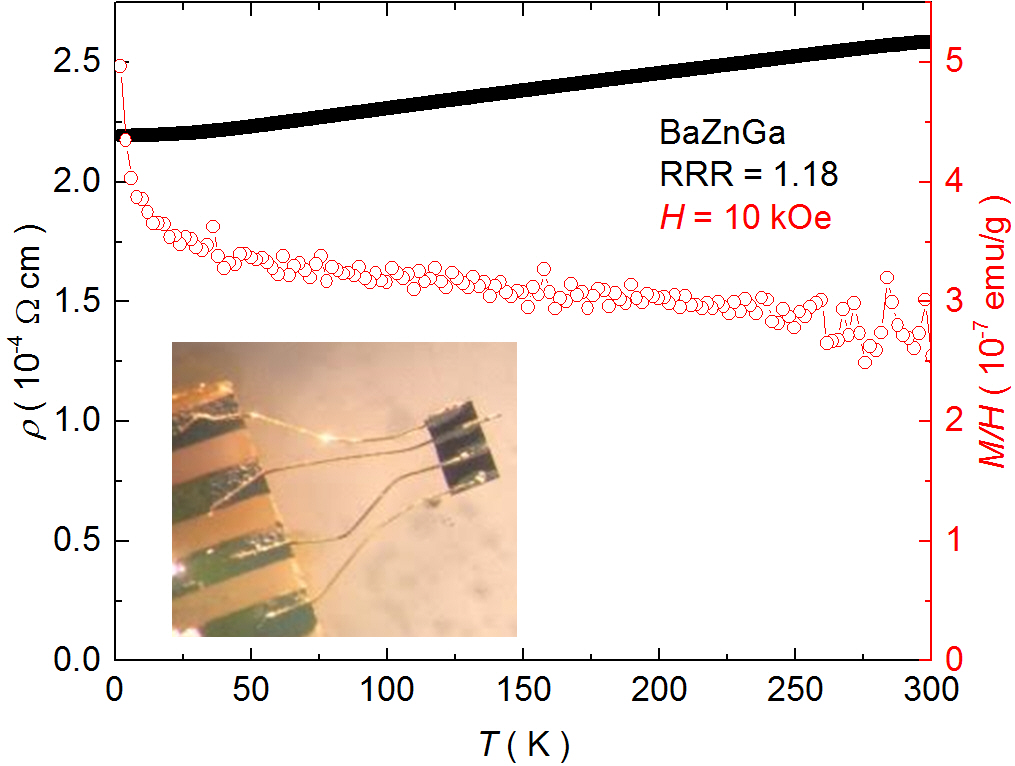}%
	\caption{(color online) Temperature dependent resistivity.(left axis, black filled circles) Temperature dependent magnetic susceptibility of BaZnGa measured at $H$\,=\,10\,kOe.(right axis, red open circle) The inset shows a picture of BaZnGa with standard 4-probe contacts.
		\label{property}}
\end{figure}
  
Figure\,\ref{property} presents temperature dependent electrical resistivity, $\rho$($T$), and magnetic susceptibility, $M(T)/H$. Not surprisingly, given that the M1 and M3, Zn/Ga sites are likely disordered, the electrical resistivity decreases only slightly from its 2.5\,$\times\,10^{-4}$\,$\ohm$\,cm room temperature value upon cooling. The overall behavior of the temperature dependent resistivity is comparable to that of BaFe$_{2}$As$_{2}$ above the structural/magnetic transition temperature, which shares the same space group for $T\,>$\,134\,K.\cite{Canfield2010} 

The magnetic susceptibility also manifests only a modest temperature dependence that can be described as weak Pauli paramagnetism with a low temperature impurity tail. This is similar in size and temperature dependence to YCo$_{2}$Zn$_{20}$\cite{Jia2007} or non-magnetic quasicrystals.\cite{Fisher1999} Neither resistivity nor susceptibility data give any indication of phase transitions between 1.8 and 300\,K. 
  
\section{Conclusion}

BaZnGa is a new ternary compound that forms in a new crystal structure (tI36). It has three, unique Ba sites and three M = Zn/Ga sites. Although there is clear chemical disorder on these sites (as supported by the poor metallic behavior of the temperature dependent electrical resistivity) density functional theory analysis suggests that the M2 site is fully occupied by Ga and the M1 and M3 sites have Zn/Ga mixing. Measurements of temperature dependent electrical resistivity and magnetic susceptibility suggest that there are no electronic, magnetic or structural phase transitions for 1.8\,K\,$<T\,<$\,300\,K.

\begin{acknowledgements}
 The authors would like to thank K. Galayda, R. S. Houk, and W. Straszheim for helpful discussions. Research are supported by the U.S. Department of Energy, Office of Basic Energy Sciences, Division of Materials Sciences and Engineering, and the grant of computer time at the National Energy Research Scientific Computing Center (NERSC) in Berkeley, CA. Ames Laboratory is operated for the U.S. Department of Energy by the Iowa State University under Contract No. DE-AC02-07CH11358. N.H.J, W.R.M and S.M. is supported by the Gordon and Betty Moore Foundation EPiQS Initiative (Grant No. GBMF4411).
\end{acknowledgements}

\section*{References}

\begin{thebibliography}{10}%
	\makeatletter
	\providecommand \@ifxundefined [1]{%
		\@ifx{#1\undefined}
	}%
	\providecommand \@ifnum [1]{%
		\ifnum #1\expandafter \@firstoftwo
		\else \expandafter \@secondoftwo
		\fi
	}%
	\providecommand \@ifx [1]{%
		\ifx #1\expandafter \@firstoftwo
		\else \expandafter \@secondoftwo
		\fi
	}%
	\providecommand \natexlab [1]{#1}%
	\providecommand \enquote  [1]{``#1''}%
	\providecommand \bibnamefont  [1]{#1}%
	\providecommand \bibfnamefont [1]{#1}%
	\providecommand \citenamefont [1]{#1}%
	\providecommand \href@noop [0]{\@secondoftwo}%
	\providecommand \href [0]{\begingroup \@sanitize@url \@href}%
	\providecommand \@href[1]{\@@startlink{#1}\@@href}%
	\providecommand \@@href[1]{\endgroup#1\@@endlink}%
	\providecommand \@sanitize@url [0]{\catcode `\\12\catcode `\$12\catcode
		`\&12\catcode `\#12\catcode `\^12\catcode `\_12\catcode `\%12\relax}%
	\providecommand \@@startlink[1]{}%
	\providecommand \@@endlink[0]{}%
	\providecommand \url  [0]{\begingroup\@sanitize@url \@url }%
	\providecommand \@url [1]{\endgroup\@href {#1}{\urlprefix }}%
	\providecommand \urlprefix  [0]{URL }%
	\providecommand \Eprint [0]{\href }%
	\providecommand \doibase [0]{http://dx.doi.org/}%
	\providecommand \selectlanguage [0]{\@gobble}%
	\providecommand \bibinfo  [0]{\@secondoftwo}%
	\providecommand \bibfield  [0]{\@secondoftwo}%
	\providecommand \translation [1]{[#1]}%
	\providecommand \BibitemOpen [0]{}%
	\providecommand \bibitemStop [0]{}%
	\providecommand \bibitemNoStop [0]{.\EOS\space}%
	\providecommand \EOS [0]{\spacefactor3000\relax}%
	\providecommand \BibitemShut  [1]{\csname bibitem#1\endcsname}%
	\let\auto@bib@innerbib\@empty
	\bibitem [{\citenamefont {Mun}\ \emph {et~al.}(2012)\citenamefont {Mun},
		\citenamefont {Ko}, \citenamefont {Miller}, \citenamefont {Samolyuk},
		\citenamefont {Bud'ko},\ and\ \citenamefont {Canfield}}]{Mun2012}%
	\BibitemOpen
	\bibfield  {author} {\bibinfo {author} {\bibfnamefont {E.}~\bibnamefont
			{Mun}}, \bibinfo {author} {\bibfnamefont {H.}~\bibnamefont {Ko}}, \bibinfo
		{author} {\bibfnamefont {G.~J.}\ \bibnamefont {Miller}}, \bibinfo {author}
		{\bibfnamefont {G.~D.}\ \bibnamefont {Samolyuk}}, \bibinfo {author}
		{\bibfnamefont {S.~L.}\ \bibnamefont {Bud'ko}}, \ and\ \bibinfo {author}
		{\bibfnamefont {P.~C.}\ \bibnamefont {Canfield}},\ }\href {\doibase
		10.1103/PhysRevB.85.035135} {\bibfield  {journal} {\bibinfo  {journal} {Phys.
				Rev. B}\ }\textbf {\bibinfo {volume} {85}},\ \bibinfo {pages} {035135}
		(\bibinfo {year} {2012})}\BibitemShut {NoStop}%
	\bibitem [{\citenamefont {Wu}\ \emph {et~al.}(2016)\citenamefont {Wu},
		\citenamefont {Wang}, \citenamefont {Mun}, \citenamefont {Johnson},
		\citenamefont {Mou}, \citenamefont {Huang}, \citenamefont {Lee},
		\citenamefont {Bud'ko}, \citenamefont {Canfield},\ and\ \citenamefont
		{Kaminski}}]{WuPtSn2016}%
	\BibitemOpen
	\bibfield  {author} {\bibinfo {author} {\bibfnamefont {Y.}~\bibnamefont
			{Wu}}, \bibinfo {author} {\bibfnamefont {L.-L.}\ \bibnamefont {Wang}},
		\bibinfo {author} {\bibfnamefont {E.}~\bibnamefont {Mun}}, \bibinfo {author}
		{\bibfnamefont {D.~D.}\ \bibnamefont {Johnson}}, \bibinfo {author}
		{\bibfnamefont {D.}~\bibnamefont {Mou}}, \bibinfo {author} {\bibfnamefont
			{L.}~\bibnamefont {Huang}}, \bibinfo {author} {\bibfnamefont
			{Y.}~\bibnamefont {Lee}}, \bibinfo {author} {\bibfnamefont {S.~L.}\
			\bibnamefont {Bud'ko}}, \bibinfo {author} {\bibfnamefont {P.~C.}\
			\bibnamefont {Canfield}}, \ and\ \bibinfo {author} {\bibfnamefont
			{A.}~\bibnamefont {Kaminski}},\ }\href@noop {} {\bibfield  {journal}
		{\bibinfo  {journal} {Nat. Phys.}\ }\textbf {\bibinfo {volume} {12}},\
		\bibinfo {pages} {667} (\bibinfo {year} {2016})}\BibitemShut {NoStop}%
	\bibitem [{\citenamefont {Canfield}\ \emph {et~al.}(2010)\citenamefont
		{Canfield}, \citenamefont {Caudle}, \citenamefont {Ho}, \citenamefont
		{Kreyssig}, \citenamefont {Nandi}, \citenamefont {Kim}, \citenamefont {Lin},
		\citenamefont {Kracher}, \citenamefont {Dennis}, \citenamefont {McCallum},\
		and\ \citenamefont {Goldman}}]{Canfield20102}%
	\BibitemOpen
	\bibfield  {author} {\bibinfo {author} {\bibfnamefont {P.~C.}\ \bibnamefont
			{Canfield}}, \bibinfo {author} {\bibfnamefont {M.~L.}\ \bibnamefont
			{Caudle}}, \bibinfo {author} {\bibfnamefont {C.-S.}\ \bibnamefont {Ho}},
		\bibinfo {author} {\bibfnamefont {A.}~\bibnamefont {Kreyssig}}, \bibinfo
		{author} {\bibfnamefont {S.}~\bibnamefont {Nandi}}, \bibinfo {author}
		{\bibfnamefont {M.~G.}\ \bibnamefont {Kim}}, \bibinfo {author} {\bibfnamefont
			{X.}~\bibnamefont {Lin}}, \bibinfo {author} {\bibfnamefont {A.}~\bibnamefont
			{Kracher}}, \bibinfo {author} {\bibfnamefont {K.~W.}\ \bibnamefont {Dennis}},
		\bibinfo {author} {\bibfnamefont {R.~W.}\ \bibnamefont {McCallum}}, \ and\
		\bibinfo {author} {\bibfnamefont {A.~I.}\ \bibnamefont {Goldman}},\ }\href
	{\doibase 10.1103/PhysRevB.81.020201} {\bibfield  {journal} {\bibinfo
			{journal} {Phys. Rev. B}\ }\textbf {\bibinfo {volume} {81}},\ \bibinfo
		{pages} {020201} (\bibinfo {year} {2010})}\BibitemShut {NoStop}%
	\bibitem [{\citenamefont {Goldman}\ \emph {et~al.}(2013)\citenamefont
		{Goldman}, \citenamefont {Kong}, \citenamefont {Kreyssig}, \citenamefont
		{Jesche}, \citenamefont {Ramazanoglu}, \citenamefont {Dennis}, \citenamefont
		{Bud’ko},\ and\ \citenamefont {Canfield}}]{Goldman2013}%
	\BibitemOpen
	\bibfield  {author} {\bibinfo {author} {\bibfnamefont {A.~I.}\ \bibnamefont
			{Goldman}}, \bibinfo {author} {\bibfnamefont {T.}~\bibnamefont {Kong}},
		\bibinfo {author} {\bibfnamefont {A.}~\bibnamefont {Kreyssig}}, \bibinfo
		{author} {\bibfnamefont {A.}~\bibnamefont {Jesche}}, \bibinfo {author}
		{\bibfnamefont {M.}~\bibnamefont {Ramazanoglu}}, \bibinfo {author}
		{\bibfnamefont {K.~W.}\ \bibnamefont {Dennis}}, \bibinfo {author}
		{\bibfnamefont {S.~L.}\ \bibnamefont {Bud'ko}}, \ and\ \bibinfo {author}
		{\bibfnamefont {P.~C.}\ \bibnamefont {Canfield}},\ }\href
	{http://dx.doi.org/10.1038/nmat3672} {\bibfield  {journal} {\bibinfo
			{journal} {Nat Mater}\ }\textbf {\bibinfo {volume} {12}},\ \bibinfo {pages}
		{714} (\bibinfo {year} {2013})}\BibitemShut {NoStop}%
	\bibitem [{\citenamefont {Jia}\ \emph {et~al.}(2007)\citenamefont {Jia},
		\citenamefont {Bud/'ko}, \citenamefont {Samolyuk},\ and\ \citenamefont
		{Canfield}}]{Jia20072}%
	\BibitemOpen
	\bibfield  {author} {\bibinfo {author} {\bibfnamefont {S.}~\bibnamefont
			{Jia}}, \bibinfo {author} {\bibfnamefont {S.~L.}\ \bibnamefont {Bud'ko}},
		\bibinfo {author} {\bibfnamefont {G.~D.}\ \bibnamefont {Samolyuk}}, \ and\
		\bibinfo {author} {\bibfnamefont {P.~C.}\ \bibnamefont {Canfield}},\ }\href
	{http://dx.doi.org/10.1038/nphys568} {\bibfield  {journal} {\bibinfo
			{journal} {Nat Phys}\ }\textbf {\bibinfo {volume} {3}},\ \bibinfo {pages}
		{334} (\bibinfo {year} {2007})}\BibitemShut {NoStop}%
	\bibitem [{\citenamefont {Torikachvili}\ \emph {et~al.}(2007)\citenamefont
		{Torikachvili}, \citenamefont {Jia}, \citenamefont {Mun}, \citenamefont
		{Hannahs}, \citenamefont {Black}, \citenamefont {Neils}, \citenamefont
		{Martien}, \citenamefont {Bud'Ko},\ and\ \citenamefont
		{Canfield}}]{torikachvili2007}%
	\BibitemOpen
	\bibfield  {author} {\bibinfo {author} {\bibfnamefont {M.}~\bibnamefont
			{Torikachvili}}, \bibinfo {author} {\bibfnamefont {S.}~\bibnamefont {Jia}},
		\bibinfo {author} {\bibfnamefont {E.}~\bibnamefont {Mun}}, \bibinfo {author}
		{\bibfnamefont {S.}~\bibnamefont {Hannahs}}, \bibinfo {author} {\bibfnamefont
			{R.}~\bibnamefont {Black}}, \bibinfo {author} {\bibfnamefont
			{W.}~\bibnamefont {Neils}}, \bibinfo {author} {\bibfnamefont
			{D.}~\bibnamefont {Martien}}, \bibinfo {author} {\bibfnamefont
			{S.}~\bibnamefont {Bud'Ko}}, \ and\ \bibinfo {author} {\bibfnamefont
			{P.}~\bibnamefont {Canfield}},\ }\href@noop {} {\bibfield  {journal}
		{\bibinfo  {journal} {Proc. Natl. Acad. Sci. U.S.A.}\
		}\textbf {\bibinfo {volume} {104}},\ \bibinfo {pages} {9960} (\bibinfo {year}
		{2007})}\BibitemShut {NoStop}%
	\bibitem [{\citenamefont {Jesche}\ \emph {et~al.}(2014)\citenamefont {Jesche},
		\citenamefont {McCallum}, \citenamefont {Thimmaiah}, \citenamefont {Jacobs},
		\citenamefont {Taufour}, \citenamefont {Kreyssig}, \citenamefont {Houk},
		\citenamefont {Bud’ko},\ and\ \citenamefont {Canfield}}]{Jesche2014}%
	\BibitemOpen
	\bibfield  {author} {\bibinfo {author} {\bibfnamefont {A.}~\bibnamefont
			{Jesche}}, \bibinfo {author} {\bibfnamefont {R.~W.}\ \bibnamefont
			{McCallum}}, \bibinfo {author} {\bibfnamefont {S.}~\bibnamefont {Thimmaiah}},
		\bibinfo {author} {\bibfnamefont {J.~L.}\ \bibnamefont {Jacobs}}, \bibinfo
		{author} {\bibfnamefont {V.}~\bibnamefont {Taufour}}, \bibinfo {author}
		{\bibfnamefont {A.}~\bibnamefont {Kreyssig}}, \bibinfo {author}
		{\bibfnamefont {R.~S.}\ \bibnamefont {Houk}}, \bibinfo {author}
		{\bibfnamefont {S.~L.}\ \bibnamefont {Bud’ko}}, \ and\ \bibinfo {author}
		{\bibfnamefont {P.~C.}\ \bibnamefont {Canfield}},\ }\href
	{http://dx.doi.org/10.1038/ncomms4333} {\bibfield  {journal} {\bibinfo
			{journal} {Nat. Comm.}\ }\textbf {\bibinfo {volume} {5}},\ \bibinfo {pages}
		{3333} (\bibinfo {year} {2014})}\BibitemShut {NoStop}%
	\bibitem [{\citenamefont {Okamoto}(2000)}]{ASM2000}%
	\BibitemOpen
	\bibfield  {author} {\bibinfo {author} {\bibfnamefont {H.}~\bibnamefont
			{Okamoto}},\ }\href@noop {} {\emph {\bibinfo {title} {Phase Diagrams for
				Binary Alloys, Desk Handbook}}}\ (\bibinfo  {publisher} {ASM International,
		Materials Park, OH},\ \bibinfo {year} {2000})\BibitemShut {NoStop}%
	\bibitem [{\citenamefont {Canfield}\ and\ \citenamefont
		{Fisk}(1992)}]{Canfield1992}%
	\BibitemOpen
	\bibfield  {author} {\bibinfo {author} {\bibfnamefont {P.~C.}\ \bibnamefont
			{Canfield}}\ and\ \bibinfo {author} {\bibfnamefont {Z.}~\bibnamefont
			{Fisk}},\ }\href@noop {} {\bibfield  {journal} {\bibinfo  {journal} {Philos.
				Mag. B}\ }\textbf {\bibinfo {volume} {65}},\ \bibinfo {pages} {1117}
		(\bibinfo {year} {1992})}\BibitemShut {NoStop}%
	\bibitem [{\citenamefont {Canfield}\ \emph {et~al.}(2016)\citenamefont
		{Canfield}, \citenamefont {Kong}, \citenamefont {Kaluarachchi},\ and\
		\citenamefont {Jo}}]{Canfield2016}%
	\BibitemOpen
	\bibfield  {author} {\bibinfo {author} {\bibfnamefont {P.~C.}\ \bibnamefont
			{Canfield}}, \bibinfo {author} {\bibfnamefont {T.}~\bibnamefont {Kong}},
		\bibinfo {author} {\bibfnamefont {U.~S.}\ \bibnamefont {Kaluarachchi}}, \
		and\ \bibinfo {author} {\bibfnamefont {N.~H.}\ \bibnamefont {Jo}},\
	}\href@noop {} {\bibfield  {journal} {\bibinfo  {journal} {Philos. Mag.}\
	}\textbf {\bibinfo {volume} {96}},\ \bibinfo {pages} {84} (\bibinfo {year}
	{2016})}\BibitemShut {NoStop}%
\bibitem {Bruker2002}%
\textit{SMART for Windows NT/2000. Version 6.148.}, Bruker AXS Inc., M. Wisconsin, USA 2002.

\bibitem {Bruker2013}%
\newblock{\textit{SAINT plus. Version 8.30.}}, Bruker AXS Inc., M. Wisconsin,
USA, 2013

\bibitem {Bruker2000}%
\newblock{\textit{SHELXTL, Bruker Analytical X-ray
		Systems}}, Bruker Inc., Madison, WI, USA, 6th ed.

\bibitem {Palatinus2007}%
L.~Palatinus and G.~Chapuis
\newblock{J. Appl. Crystallogr}, \textbf{40}, 786, 2007	

\bibitem {Kohn1965}%
W.~Kohn and L.~J.~Sham
\newblock{Phys. Rev.}, \textbf{140}, A1133, 1965.

\bibitem {Kresse1996}%
G.~Kresse and J.~Furthm\"uller
\newblock{Phys. Rev. B}, \textbf{54}, 11169, 1996.

\bibitem {Blochl1994}%
P.~E.~Bl\"ochl
\newblock{Phys. Rev. B}, \textbf{50}, 17953, 1994.

\bibitem {Kresse1999}%
G.~Kresse and D.~Joubert
\newblock{Phys. Rev. B}, \textbf{59}, 1758, 1999.

\bibitem {Perdew1996}%
J.~P.~Perdew, K.~Burke, M.~Ernzerhof
\newblock{Phys. Rev. Lett.}, \textbf{77}, 3865, 1996.

\bibitem {Monkhorst1976}%
H.~J.~Monkhorst and J.~D.~Pack
\newblock{Phys. Rev. B.}, \textbf{13}, 5188, 1976.

\bibitem {Canfield2010}%
P.~C.~Canfield and S.~L.~Bud'ko
\newblock{Annu. Rev. Condens. Matter Phys.}, \textbf{1}, 27, 2010.

\bibitem {Jia2007}%
S. Jia, S.~L.~Bud'ko, G.~D.~Samolyuk, P.~C.~Canfield 
\newblock{Nat. Phys.}, \textbf{3}, 334, 2007.

\bibitem {Fisher1999}%
I.~R.~Fisher, K.~O.~Choen, A.~F.~Panchula, P.~C.~Canfield, M.~Chernikov, H.~R.~Ott, K.~Dnnis 
\newblock{Phys. Rev. B}, \textbf{59}, 1, 308, 1999.		

\end{thebibliography}
%

\end{document}